\documentstyle[12pt,aasms4]{article}
 
\slugcomment{To appear in the Astrophysical Journal
(letters)}
 
\begin{document}
 
\title{Detection of HI 21cm Absorption by the Warm Neutral Medium}
 
\author{C.L. Carilli}
\affil{National Radio Astronomy Observatory, P.O. Box O, Socorro, NM,
87801 \\} 
\author{K.S. Dwarakanath}
\affil{Raman Research Institute, Bangalore 560 080, India \\} 
\author{W.M. Goss}
\affil{National Radio Astronomy Observatory, P.O. Box O, Socorro, NM,
87801 \\} 

\begin{abstract}

We have detected HI  21cm line absorption by the Warm Neutral Medium
(WNM) using the Westerbork Synthesis Radio Telescope. The absorption
was detected toward Cygnus A at LSR velocities of --40 and --70 km
s$^{-1}$. These two velocity ranges were previously identified as
being relatively free of cold absorbing clouds. The measured  optical
depth for the WNM along the line of sight to Cygnus A 
is 8.9$\pm$1.9 $\times$10$^{-4}$ at --70 km s$^{-1}$,
and 8.5$\pm$2.0 $\times$10$^{-4}$ at --40 km s$^{-1}$, with
corresponding spin temperatures of 6000$\pm$1700 K and 
4800$\pm$1600 K, respectively. The volume filling factor for the WNM
appears to be fairly high (f $\approx$ 0.4). 

{\sl Subject headings: ISM: structure and atoms -- 
Galaxy: fundamental parameters -- Radio Lines: ISM}

\end{abstract}

\section {Introduction}

Early observations of the 21cm line of neutral hydrogen in emission
and absorption in the Galaxy showed that toward 
any given direction the HI line seen in emission was always broader than the
corresponding HI absorption line. In addition, most lines of sight showed a 
broad HI emission line with no corresponding HI absorption 
(Clark, Radhakrishnan, \&  Wilson 1962, Clark 1965, Radhakrishnan et
al. 1972). These observations 
led to the proposal of a `raisin-pudding' model of the
interstellar medium  (Field, Goldsmith, \& Habing 1969), 
in which dense cool HI clouds at 10$^2$ K (the Cold Neutral
Medium; CNM) are enveloped by a  warmer, lower density gas
at 10$^{4}$ K (the Warm Neutral Medium; WNM). 
More recent models for the interstellar medium include
a high filling factor, Hot Ionized Medium (HIM) at 10$^6$ K, 
resulting from supernovae explosions (McKee and Ostriker 1977).
The phases of the interstellar medium are thought to be in pressure
equilibrium, although the
relative filling factors of the different phases remain uncertain,
and the filling factors may vary with position in galaxies
(Brinks 1990, Braun and Walterbos 1992).

Kulkarni and Heiles (1988) point out that of all the phases of the
interstellar medium (ISM), the WNM remains the least well
understood. Although HI 21cm emission from the WNM is detected
easily in every direction in the sky,  an accurate value for  the
temperature of the WNM is lacking.
The standard  method with which to  determine the 
temperature of interstellar HI
is to compare HI  column densities determined by observation of the
21cm line in emission in a given region, with those determined by
observation of 21cm absorption towards background continuum sources
(Clark 1965). The result is the excitation, or
spin temperature, of the gas. The spin temperature, T$_s$, is thought
to be close to the kinetic temperature in most Galactic environments. 
There are three  well known problems
with this method.  The first is that one has to
interpolate HI column densities determined  off-source to 
the position of the continuum source. For the WNM
this is not a severe problem, since the column density distribution
appears to be fairly smooth on scales $\ge$ 10$'$ 
(Kalberla, Schwarz, and Goss 1985). Second,
when  looking for absorption by the WNM, there is the 
confusion caused by the plethora of absorption lines arising in
the CNM.  And third, there is the problem of
having both emission and absorption in the on-source beam,  and
the  related problem of  `stray-radiation', or emission coming  in
through the sidelobes of the telescope beam.  
This is especially a problem for single dish observations, for which
beam sizes  are typically much larger than the background continuum
sources. 

Kulkarni and Heiles (1988) review measurements of the temperature of
the WNM. They conclude that the current data, which entails mostly
lower limits to T$_s$, are consistent with a temperature range of 5000
K to 8000 K for most of the WNM. They emphasize that the
current data are sparse, and that `observational confirmation of this
conclusion is {\sl crucial}.' 
The single existing experiment claiming detection of absorption by the
WNM is that by Mebold and Hills (1975).  They studied  HI emission and
absorption  in the vicinity of  the bright extragalactic radio source
Cygnus A using the  Effelsberg 100 m telescope.
Their experiment consisted of eight off-source pointings encircling
the continuum source, to determine the HI column density
distribution in the 
region, and  on-source observations to look for absorption.
They found two  inter-arm velocity ranges with no narrow
absorption features due to the CNM.  At these
velocities their on-source spectrum of Cygnus A shows 
HI in {\sl  emission}, however, the amount of emission
seen on-source is less than that expected from the interpolation
of the off-source pointings to the source position. 
From this they {\sl infer} absorption 
towards the continuum source.  Mebold and Hills
estimate an optical depth for the WNM between 
6$\times$10$^{-4}$ and 11$\times$10$^{-4}$,
and a spin temperature between 6400 K and 7800 K.  
Kalberla, Mebold, and Reich (1980) re-analyzed these data, 
correcting for stray radiation, and  reduced the spin temperature
estimates by about 20$\%$. 

In this letter we present a direct verification of absorption 
toward Cygnus A in the CNM-free velocity ranges using the Westerbork
Synthesis Radio Telescope (WSRT). We then derive the spin temperature for
the WNM in the two CNM-free velocity ranges using the WSRT absorption 
measurements and  emission spectra from the Effelsberg and 
Dwingeloo telescopes. We discuss these
results in terms of standard equilibrium models for the two-phase
neutral atomic ISM. 
  
\section{Observations of CNM-free Regions toward Cygnus A}

The CNM-free LSR velocity ranges toward Cygnus A are
from --40 to --36 km s$^{-1}$, and --63 to --78 km s$^{-1}$, corresponding
to inter-arm velocities,  between the Orion (0 km s$^{-1}$), 
Perseus (--50  km s$^{-1}$), and Outer (-85 km s$^{-1}$) spiral arms (Mebold
and Hills 1975). We have searched for absorption by the WNM at these
inter-arm velocities using the  WSRT. 
Using an interferometer mitigates problems of confusion by
stray radiation and in-beam emission. 
Observations were made on May 24 and 25, 1996, July 15 and 16, 1996, 
June 28 and 29, 1997,
and August 22 and 23, 1997, for 12 hours each day. The total bandwidth
was 1.25 MHz with 255 spectral channels 
after Hanning smoothing, corresponding to 1 km s$^{-1}$ velocity
resolution, centered at an LSR velocity of --45 km s$^{-1}$.  Absolute
flux calibration and initial phase calibration were performed using 3C
286, and residual phase errors were corrected through
self-calibration using a continuum data set made from line-free channels.

An accurate bandpass was determined using frequency switching to
`high' and `low' frequencies by $\pm$
1.25 MHz on Cygnus A itself. The cycle time for bandpass calibration
was two hours, with a 50$\%$ duty cycle, and the bandpass for each
cycle was the average of the high and low frequency settings. 
An estimate of the accuracy of the bandpass calibration process was
obtained by applying the `high' frequency bandpass solutions to the `low'
frequency spectra. The residual fractional deviations were generally $\le$
1.7$\times$10$^{-4}$. We adopt this value as a conservative 
estimate of optical depth errors introduced by the bandpass
calibration process.  


The visibility data from each day's observation were reduced
independently using the Astronomical Image Processing System (AIPS). 
The continuum emission was
subtracted using the AIPS task UVLIN, which performs a 
linear fit to the line-free channels for each visibility, and the
data from the eight days were then combined. Spectral line image `cubes'
were constructed from the combined data, and the point response
function of the array deconvolved, using the AIPS task IMAGR. The
Gaussian restoring CLEAN beam  had a FWHM = 30$''$$\times$15$''$ with
a major axis oriented north-south. 

The continuum image of Cygnus A made from line-free channels is shown in
Figure 1. The observed total flux density at 1.4 GHz for Cygnus A was
within 2$\%$ of the expected value of 1575 Jy (Baars et al. 1977).  
The source is characterized by two `hot spots' of emission at
the source extremities, separated by 2$'$, with a low frequency
`bridge' of emission connecting the two hot spots. In the following
section we analyze spectra taken at the 
the Northwest (NW) and Southeast (SE) hot spots. 
The peak surface brightness is 346 Jy beam$^{-1}$ for the NW hot spot
and 398 Jy beam$^{-1}$ for the SE hot spot. 

The HI absorption spectra toward the SE and NW peaks of Cygnus A are
shown in Figure 2. The spectra show deep, 
narrow lines due to the CNM, clustered around the spiral arm
velocities (0, --50,  and --85 km s$^{-1}$), with optical depths
approaching 0.5. There are significant variations in
optical depth for these narrow lines between the two lines of sight
separated by 2$'$. Small scale structure in the CNM is a well studied
phenomenon (Dieter et al. 1976, Diamond et al. 1989, 
Deshpande et al. 1992, Frail et al. 1994, Davis, Diamond, and
Goss 1996), and a detailed analysis of the CNM structure 
toward Cygnus A will be presented elsewhere. The root mean square
(rms) noise per channel in these spectra is 0.1 mJy channel$^{-1}$,
corresponding to an rms optical depth ($\sigma$) of about
2.5$\times$10$^{-4}$ per channel. 

Note the lack of deep, narrow absorption features in the inter-arm
velocity ranges from --40 to --36 km s$^{-1}$ (region `B' in the
spectra), and --63 to --78 km s$^{-1}$ (region `A'), as first 
pointed out by Mebold and Hills (1975). Figures 3 \& 4 
show these same spectra with an expanded flux and optical depth 
scale in order to investigate possible absorption in these two
CNM-free velocity ranges.  The observed (continuum subtracted)
surface brightnesses in regions A and B 
appear to be significantly below zero in all the spectra. The mean
optical depth values in regions A and B are listed in Table 1. 
The errors are a (quadratic) sum of the sensitivity error (ie. the rms
in off-line channels after averaging over the specified number of
channels), and the estimated error due to bandpass  calibration.  


The optical depth in velocity region B is marginally
higher  toward the SE source 
relative to the NW source (1.8$\sigma$). 
However the SE source spectrum shows a
narrow absorption line by the CNM at --30 km s$^{-1}$ which is not seen
toward the NW source. The wing of this line might perturb the
measured value in region B. Similarly, the 
spectrum of the SE component in velocity region A
shows what might be a narrow, weak absorption component at
--69 km s$^{-1}$. In the analysis below we use the optical depths in
regions A and B averaged over both SE and NW spectra (row three in
Table 1), with errors as dictated by each measurement. 

We are confident that the finite optical depths observed at the CNM-free
velocities are reliable for a number of reasons. First, the observed
optical depths  are significantly higher than the estimated 
accuracy of the bandpass calibration. Second, we have analyzed spectra
from data taken from the four epochs separately (May and July 1996, and
June and August 1997), and obtain
consistent results for each epoch. And third, we have fit Gaussian 
models to the strong absorption lines nearest in velocity to the
CNM-free regions, and we find that the finite optical depths seen in
these regions are not due to the wings of the strong lines.  We
should emphasize that we cannot rule-out a model in which the 
finite optical depth seen at the CNM-free velocities toward Cygnus A
are due to a `continuum' of low optical depth cold clouds. If this
were the case, the optical depths quoted in Table 1 would be strictly
upper limits to the true optical depth of the WNM, and the spin
temperature values quoted below then become lower limits. 

\section{Discussion}

We can combine our measurement of the HI 21cm optical depth of the WNM
with the emission measurements of Mebold and Hills (1975) to obtain a
lower limit to the spin  temperature of the WNM, using the low-optical
depth approximation: T$_B$ = $\tau$T$_s$, where $\tau$ is
the optical depth  
to 21cm absorption, and T$_B$ is the brightness temperature of the
21cm emission in the absence of absorption. 

One measurement of the emission spectrum in the direction of Cygnus A
is the Effelsberg  100m spectrum of Mebold and Hills (1975), which 
entailed the average of eight emission profiles taken at distances of
20$'$ from Cygnus A in a symmetric star pattern centered on Cygnus A. 
The brightness temperatures in velocity ranges A and B are given in
row four of Table 1. These are the  values of Mebold and Hills 
reduced by 20\% to correct for stray radiation
(Kalberla et al. 1980). Mebold and Hills quote an error of 1 K
for these measurements.  A second measurement is from the all-sky HI
survey of Hartmann and Burton (1997), using the Dwingeloo 25m
telescope, and corrected for stray radiation. The Dwingeloo values are
given in row five of Table 1, where we have averaged four measurements
surrounding Cygnus A at distances of about 20$'$. The errors represent
the scatter in the measurements. Although the resolutions of
the two observations are different (11$'$ for Effelsberg and 30$'$ for
Dwingeloo), the T$_B$ values agree to within the errors. For the 
calculation of spin temperatures 
we use the (error weighted) mean T$_B$ values listed in row six of
Table 1. The resulting spin temperature values for velocity regions A
and B are listed in row seven of Table 1, where the errors are a
quadrature sum of errors on the brightness temperature measurement
and the optical depth measurement. 

The Galactic coordinates for Cygnus A are $l$ = 76$^o$, b = 5.8$^o$.
The implied distances to the gas in velocity ranges A and B are about
12 kpc and 9 kpc, respectively, as derived from the Galactic rotation
curve of Burton (1988). The height  of this gas from the plane, $z$,
is then: $z$ $\approx$ 1 kpc. A very rough estimate of the pathlength
through the gas can be derived by assuming that the velocity ranges
are set by Galactic rotation, leading to 0.9 kpc and 0.4 kpc for A and B,
respectively.  Then using the HI column densities calculated from
Table 1 leads to  a mean density for the WNM in velocity ranges A and
B toward Cygnus A of:~  $<n_{wnm}>$ $\approx$ 0.03 cm$^{-3}$. The
ISM pressure at $z$ = 1 kpc  is thought to be about
5$\times$10$^{-14}$ dyne cm$^{-2}$ (Kulkarni and Heiles 1988,
equ. 3.30). This pressure, coupled with the mean temperature of 5400 K
for the WNM from Table 1, leads to  a density of: n$_{wnm}$ = 0.07
cm$^{-3}$. Hence the volume filling factor, f,  for the WNM at $z$ = 1
kpc is:~ f $\equiv$ ${<n_{wnm}>}\over{n_{wnm}}$ $\approx$ 0.4. This
value of  f is consistent with the estimate of Kulkarni and Heiles
(1988) of f = 0.6 for the sum of the warm neutral and warm ionized
medium, with the dominant component being the WNM in the case where
magnetic fields and 
cosmic ray pressures are substantial. While the uncertainties in the
above calculation are significant (in particular the use of small
velocity gradients to derive pathlengths), 
the general implication is that the
volume filling factor for the WNM appears to be fairly high. 

The  HI spin temperature is thought to be driven to the kinetic
temperature of the gas through the process of resonant scattering of
ambient Ly $\alpha$ photons (Field 1958).  
This process depends critically on the local
Ly$\alpha$ photon density, which in turn depends on the ionization
rate due to cosmic rays and soft X-rays since the HI is opaque to
external Ly$\alpha$ photons. Based on the analysis of Galactic low
density regions by Deguchi and Watson (1985), 
Kulkarni and Heiles (1988) state that `the ionization rate is large
enough in the Galactic environment to make the spin temperature (T$_s$)
$\approx$ kinetic  temperature (T$_k$) for
almost any HI cloud observable in the 21 cm line.' 
The current limit to the diffuse Galactic Ly $\alpha$ photon density
in the solar neighborhood is: n$_{Ly\alpha}$  $<$
2.5$\times$10$^{-6}$ cm$^{-3}$ (Holberg 1986). This limit is sufficient to
maintain T$_s$ $\approx$ T$_k$ in the WNM if T$_k \sim$ 3000 K (Field 1958). 
If the value of n$_{Ly\alpha}$ is
considerably below 10$^{-6}$ cm$^{-3}$ in the WNM, then
T$_s$ could fall significantly below T$_k$. For example, Corbelli and
Salpeter (1993) find that temperature equilibrium between 
T$_s$ and T$_k$ may not hold in the outer disks of spiral galaxies, where
N(HI) $\le$ few $\times$10$^{19}$, if the only source of ionizing
radiation is the extragalactic  background. Until a direct measurement
of n$_{Ly\alpha}$ in the WNM becomes available, we can only assume that
T$_s$ $\approx$ T$_k$ in the disk of the Galaxy.

The most detailed analysis of the equilibrium state of neutral
hydrogen in the interstellar medium
is the work of Wolfire et al. (1995). The primary heating
mechanism is photoelectric emission from dust grains, while cooling is
dominated  by fine structure lines from heavy elements (predominantly
Carbon), and by hydrogen recombination lines in warmer regions. 
In their standard model for neutral hydrogen in the Galactic plane
they show that two stable phases
exist: (i)  the CNM with T$_k$ between 40 K and 200 K, and (ii) the WNM
with  T$_k$ between 5500 K and 8700 K,  for ISM pressures in the range of 
1 to 5 $\times$ 10$^{-13}$ dyne cm$^{-2}$.
It is interesting that the measured values of T$_s$ in the
lower pressure (ie. larger $z$)
WNM toward Cygnus A  (Table 1) are generally consistent
with these ranges for the neutral, two-phase ISM model of Wolfire et
al. (1995). Corbelli and Salpeter (1993) point out that equality
between T$_s$ and T$_k$ could be used to argue that the ionizing
radiation field for the WNM is well above the extragalactic
background, and hence that `relevant energy inputs from local sources
become necessary.'  While the current data toward Cygnus A are
consistent with the value of T$_s$ 
being close to the model value of T$_k$ 
for the WNM, more accurate values for T$_s$ and T$_B$ for the
neutral hydrogen, and  n$_{Ly\alpha}$, are needed to provide a fully
constrained physical model for the Galactic WNM.

\vskip 0.2in

The National Radio Astronomy Observatory (NRAO) is a facility
of the National Science Foundation, operated under cooperative
agreement by Associated Universities, Inc.. 
The Westerbork Synthesis Radio Telescope  is operated by the
Netherlands Foundation for Research in Astronomy with financial support
from the Netherlands Organization for Scientific Research (NWO).
We wish to thank K. Anantharamiah, W.B. Burton, A.A. Deshpande,
G. Field, D. Hartmann, C. Heiles,  S. Kulkarni, U. Mebold, 
G. Srinivasan, and the referee for useful comments on this work.

\newpage

\centerline{\bf References}

Baars, J., Genzel, R., Pauliny-Toth, I., and Witzel, A. 1977, A\&A,
61, 99.

Braun, R. and Walterbos, R. 1992 Ap.J., 386, 120

Brinks, E. 1990, in {\sl The Interstellar Medium in Galaxies},
eds. H.A. Thronson and J.M. Shull (Dordrecht: Kluwer), p. 39.

Burton, W. Butler 1988, in {\sl Galactic and
Extragalactic Radio Astronomy}, eds. G. Verschuur and K. Kellerman
(Heidelberg: Springer-Verlag), p. 295

Clark, B. G. 1965, ApJ,  142, 1398.

Clark, B. G., Radhakrishnan, V., \& Wilson, R. W. 1962, ApJ, 
135, 151. 

Corbelli, E. and Salpeter, E.E. 1993, Ap.J., 419, 94

Davis, R.J., Diamond, P.J., and Goss, W.M. 1996, MNRAS, 283, 1105

Deguchi, S. and Watson, W.D. 1985, Ap.J., 290, 578

Deshpande, A. A., McCulloch, P. M., Radhakrishnan, V., \& Anantharamaiah, 
K. R. 1992, MNRAS,  258, 19p.

Diamond, P. J., Goss, W. M., Romney, J. D., Booth, R. S., Kalberla, P. M. W.,
\& Mebold, U. 1989, ApJ,  347, 302.

Dieter, N. H., Welch, W. J., \& Romney, J. D. 1976, ApJ,  206, L113.

Field, G. B. 1958, Proc. IRE,  46, 240.

Field, G. B., Goldsmith, D. W., \& Habing, H. J. 1969, ApJ,  155, L149.

Frail, D. A., Weisberg, J. M., Cordes, J. M., \& Mathers, C. 1994, ApJ,
 436, 144.

Hartmann, Dap and Burton, W.B. 1997, {\sl Atlas of Galactic Neutral
Hydrogen} (Cambridge: Cambridge University Press).

Holberg, J.B. 1986, Ap.J., 311, 969.

Kalberla, P. M. W., Mebold, U., \& Reich, W. 1980, A\&A,  82, 275.

Kalberla, P. M. W., Schwarz, U. J., \& Goss, W. M. 1985, A\&A,  144, 27.

Kulkarni, S. R. and Heiles, C. 1988, in {\sl Galactic and
Extragalactic Radio Astronomy}, eds. G. Verschuur and K. Kellerman
(Heidelberg: Springer-Verlag), p. 95

McKee, C.F. and Ostriker, J.P. 1977, Ap.J., 218, 148.

Mebold, U., \& Hills, D. 1975, A\&A,  42, 187.

Radhakrishnan, V., Goss, W. M., Murray, J. D., \& Brooks, J. W. 1972, ApJS,
24, 49.

Wolfire, M.G., Hollenbach, D., McKee, C.F., Tielens, A.G., and Bakes,
E.L. 1995, Ap.J., 443, 152.

\clearpage
\newpage

\begin{deluxetable}{ccc}
\footnotesize
\tablecaption{ \label{tbl-1}}
\tablewidth{0pt}
\tablehead{
\colhead{~} & \colhead{A}  & \colhead{B} \nl
\colhead{~} & \colhead{-67~ to~ -76 km s$^{-1}$}   & \colhead{-36~ to~
-40 km s$^{-1}$} 
}
\startdata
Optical Depth: NW & 6.8$\pm$1.9 $\times$10$^{-4}$ &
6.0$\pm$2.0 $\times$10$^{-4}$ \nl
Optical Depth: SE & 11$\pm$1.9 $\times$10$^{-4}$ &
11$\pm$2.0 $\times$10$^{-4}$ \nl
Mean Optical Depth & 8.9$\pm$1.9 $\times$10$^{-4}$ &
8.5$\pm$2.0 $\times$10$^{-4}$ \nl
T$_B$ (Effelsberg) & 5.6 $\pm$ 1 K & 4.0 $\pm$ 1 K \nl
T$_B$ (Dwingeloo) & 4.7 $\pm$ 2 K & 4.5 $\pm$ 2 K \nl
Mean T$_B$ & 5.3 $\pm$ 1 K & 4.1 $\pm$ 1 K \nl
T$_s$ & 6000 $\pm$ 1700 K & 4800 $\pm$ 1600 K \nl
\enddata
\end{deluxetable}
 
\clearpage
\newpage

\centerline{Figure Captions}

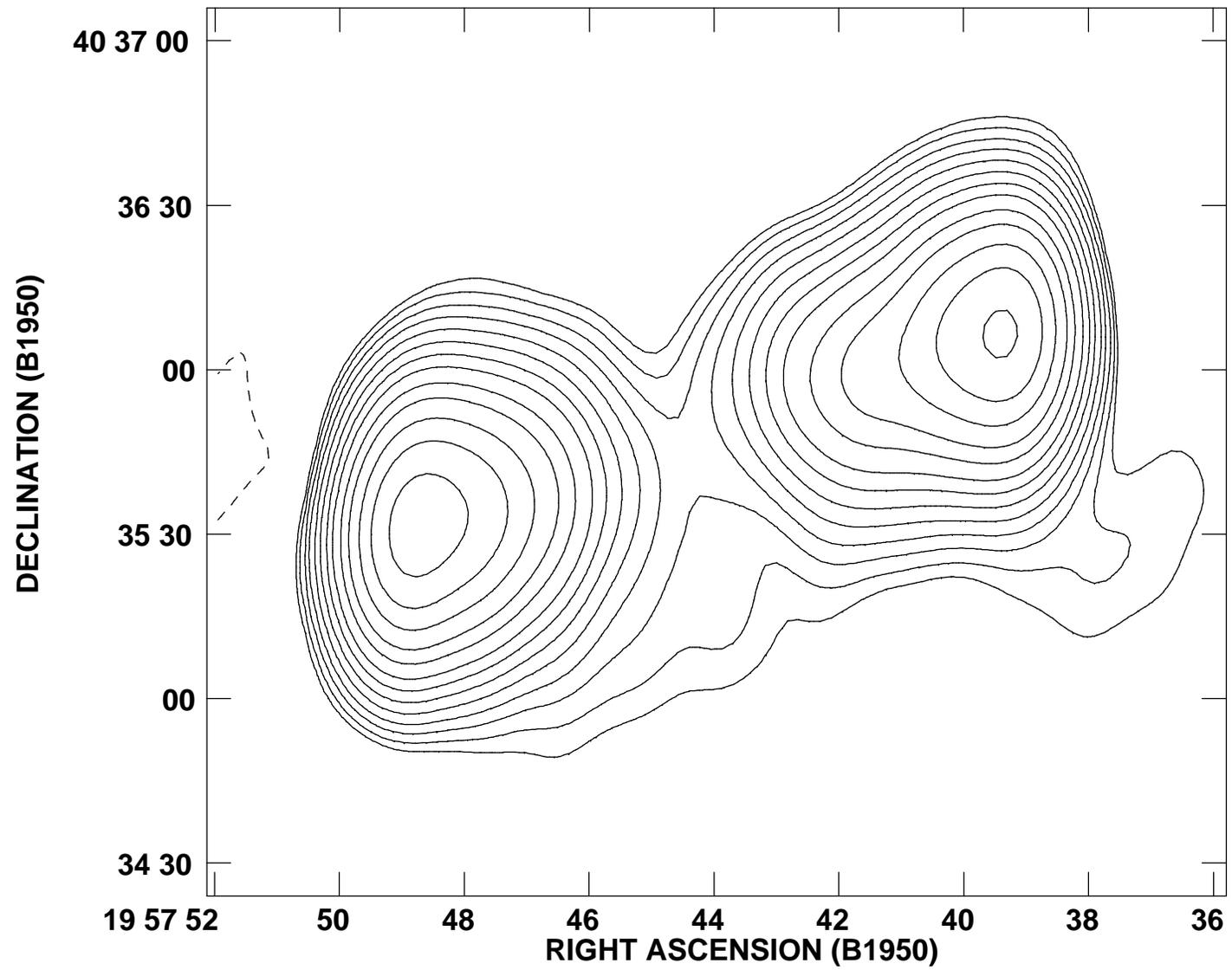
\noindent Figure 1 --
A continuum image of Cygnus A at 1.4 GHz made using line-free channels
from the WSRT observations. The total flux density is 1575 Jy. 
The contours are a geometric progression in the square root of
two. The first level is 5 Jy/beam.
Dotted contours are negative levels. The FWHM of the Gaussian
restoring beam is 30$''$$\times$15$''$ with the major axis oriented
north-south.

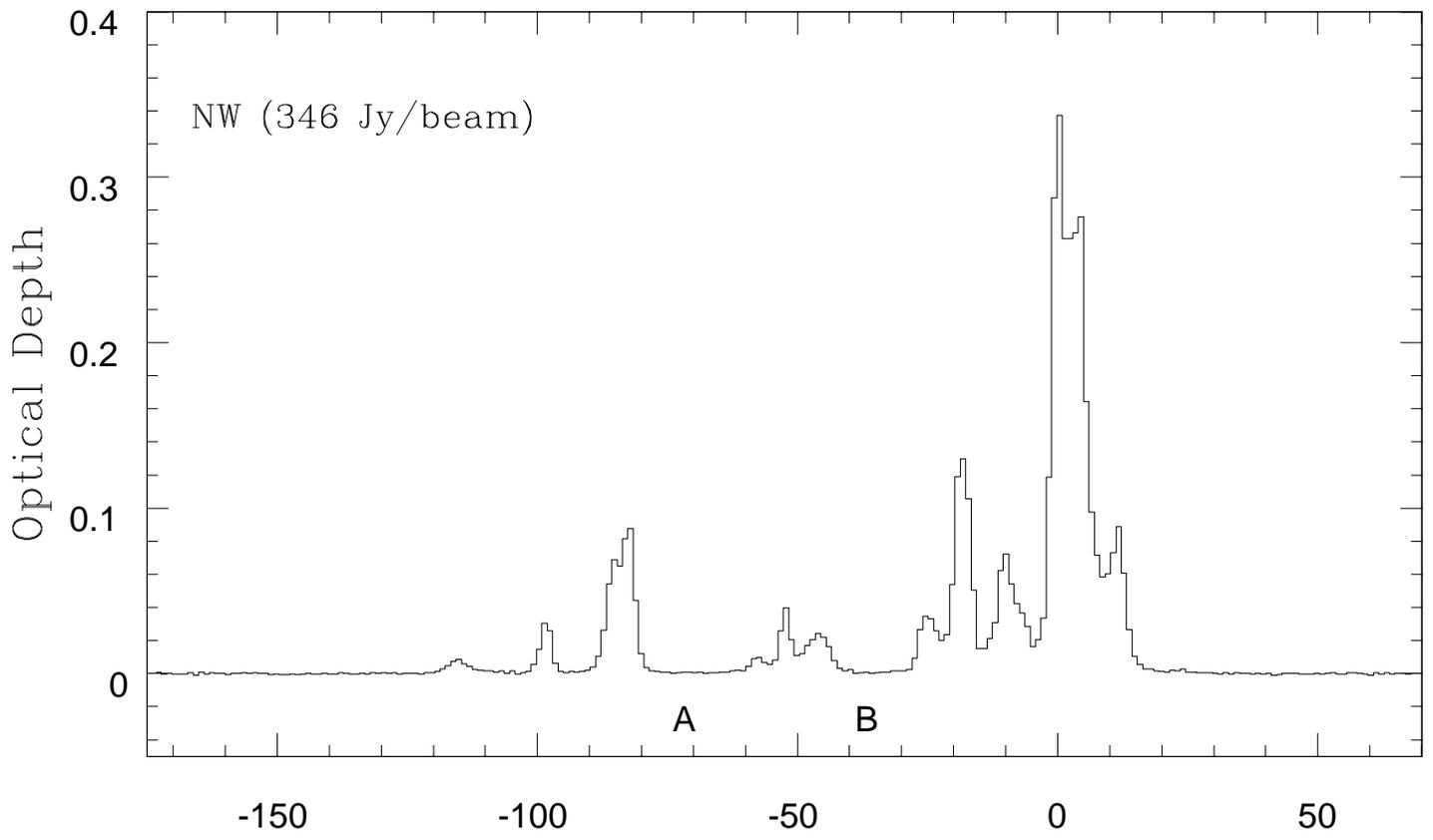
\noindent Figure 2 -- The upper frame shows the HI absorption spectrum
toward north-west peak of Cygnus A. The spectral resolution in this,
and subsequent, spectra is 1 km s$^{-1}$ channel$^{-1}$.
The spectrum has been converted
to optical depth using the continuum surface brightness of 346
Jy/beam at the resolution of Figure 1. The velocity scale is LSR. The
CNM-free velocity ranges are 
labeled `A' and `B'. The lower frame shows the corresponding spectrum
of the south-east peak, where the continuum surface brightness is 398
Jy/beam. 

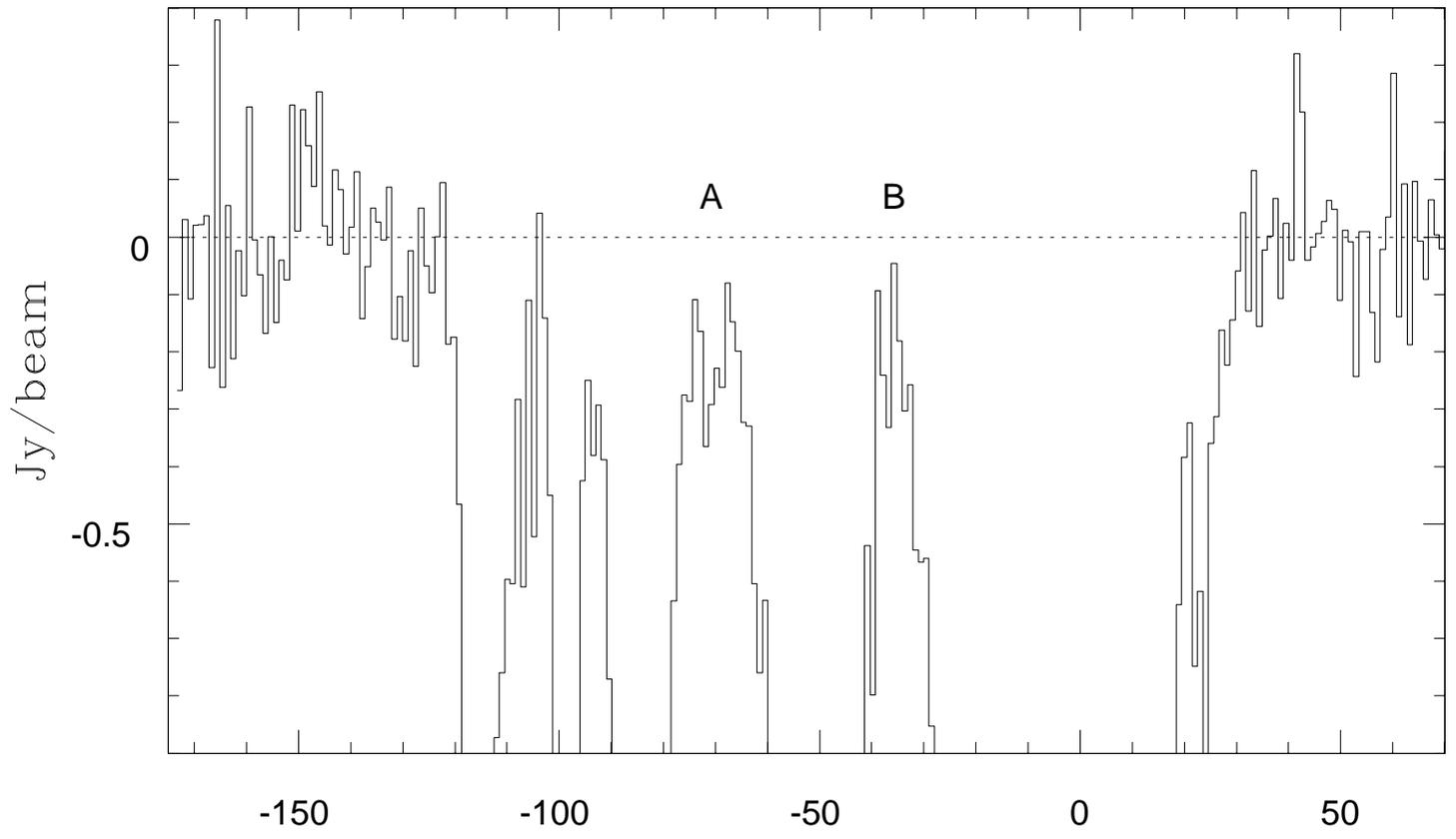
\noindent Figure 3 -- The upper frame shows the HI absorption spectrum
toward the north-west peak of Cygnus A, after subtracting the
continuum surface brightness of 346 Jy/beam. The velocity scale is
LSR. The lower frame shows the same spectrum converted to optical
depth using the continuum surface brightness of 346
Jy/beam. The optical depth scale has been expanded to show low optical
depth absorption. The CNM-free velocity ranges A and B are indicated.

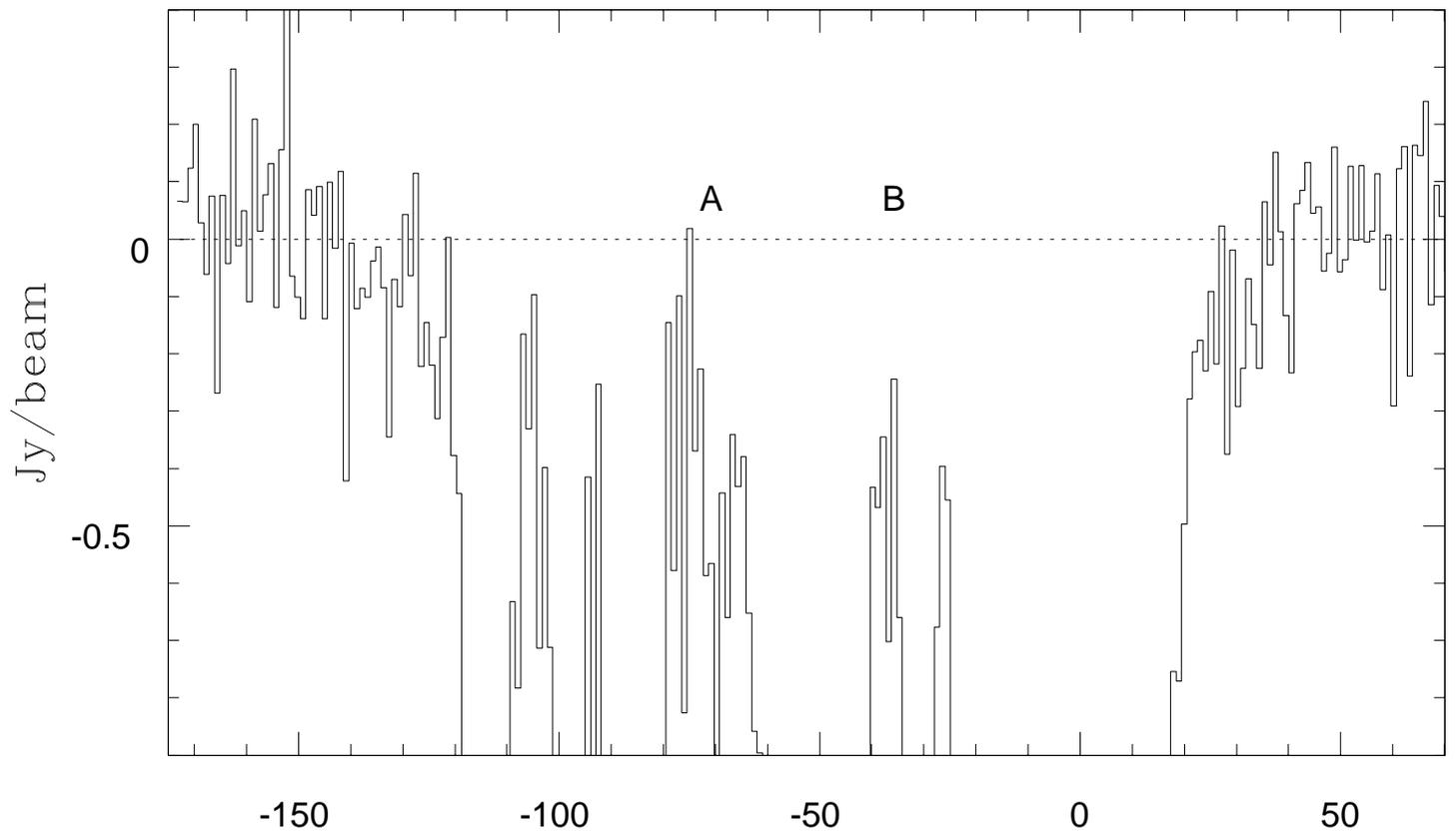
\noindent Figure 4 -- The upper frame shows the HI absorption spectrum
toward the south-east peak of Cygnus A, after subtracting the
continuum surface brightness of 398 Jy/beam. The velocity scale is
LSR. The lower frame shows the same spectrum converted to optical
depth using the continuum surface brightness of 398
Jy/beam. The optical depth scale has been expanded to show low optical
depth absorption. The CNM-free velocity ranges A and B are indicated.

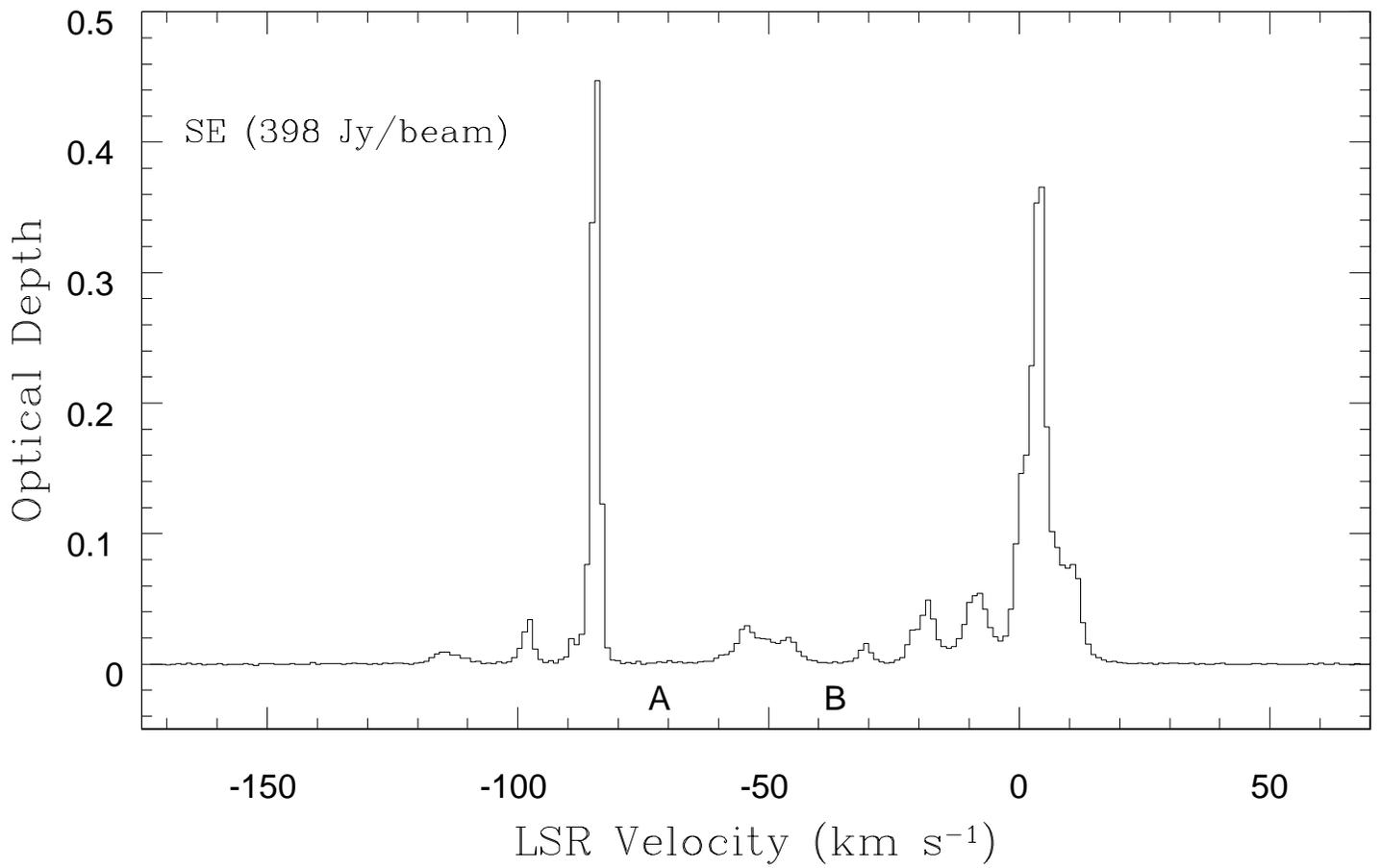
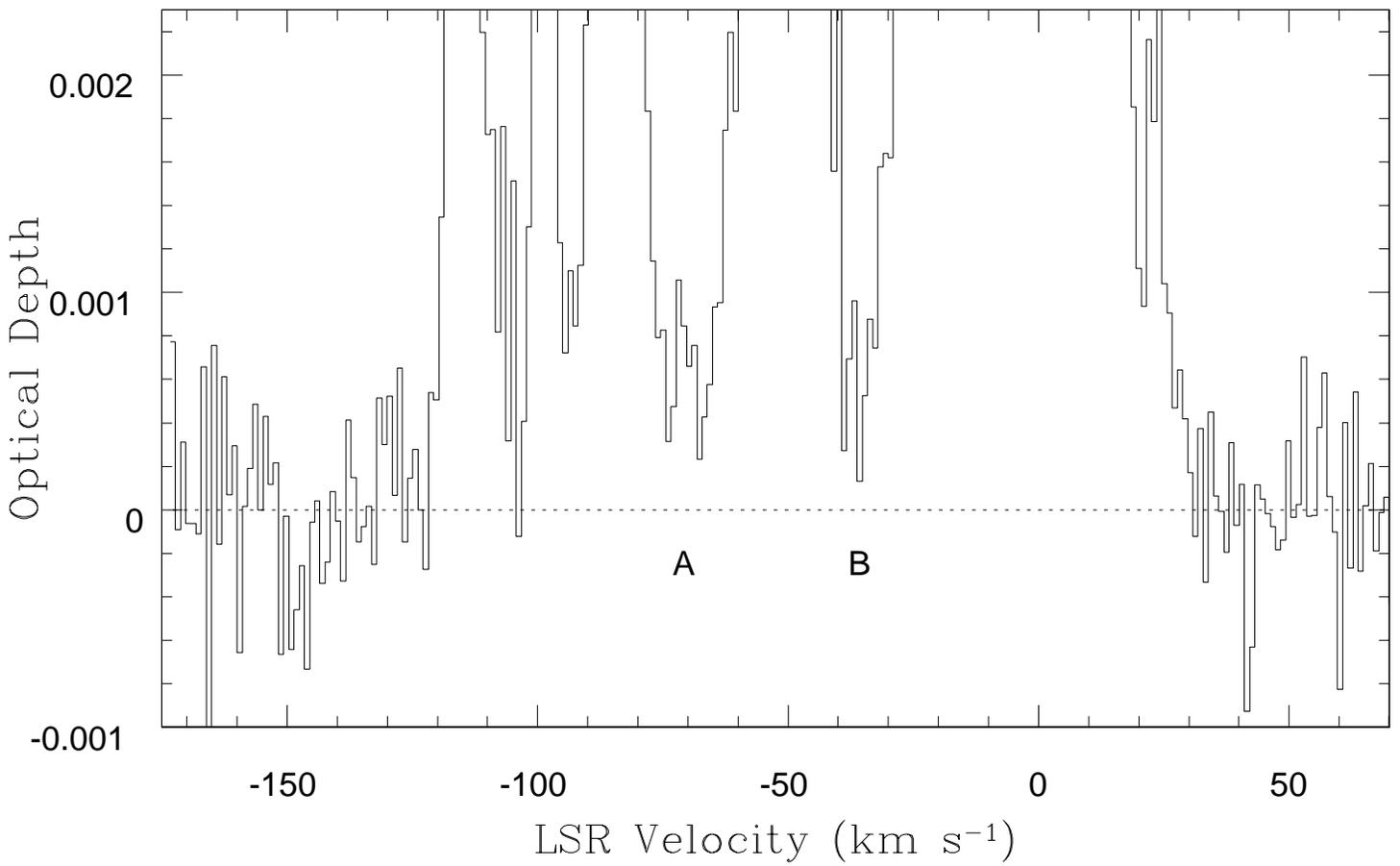
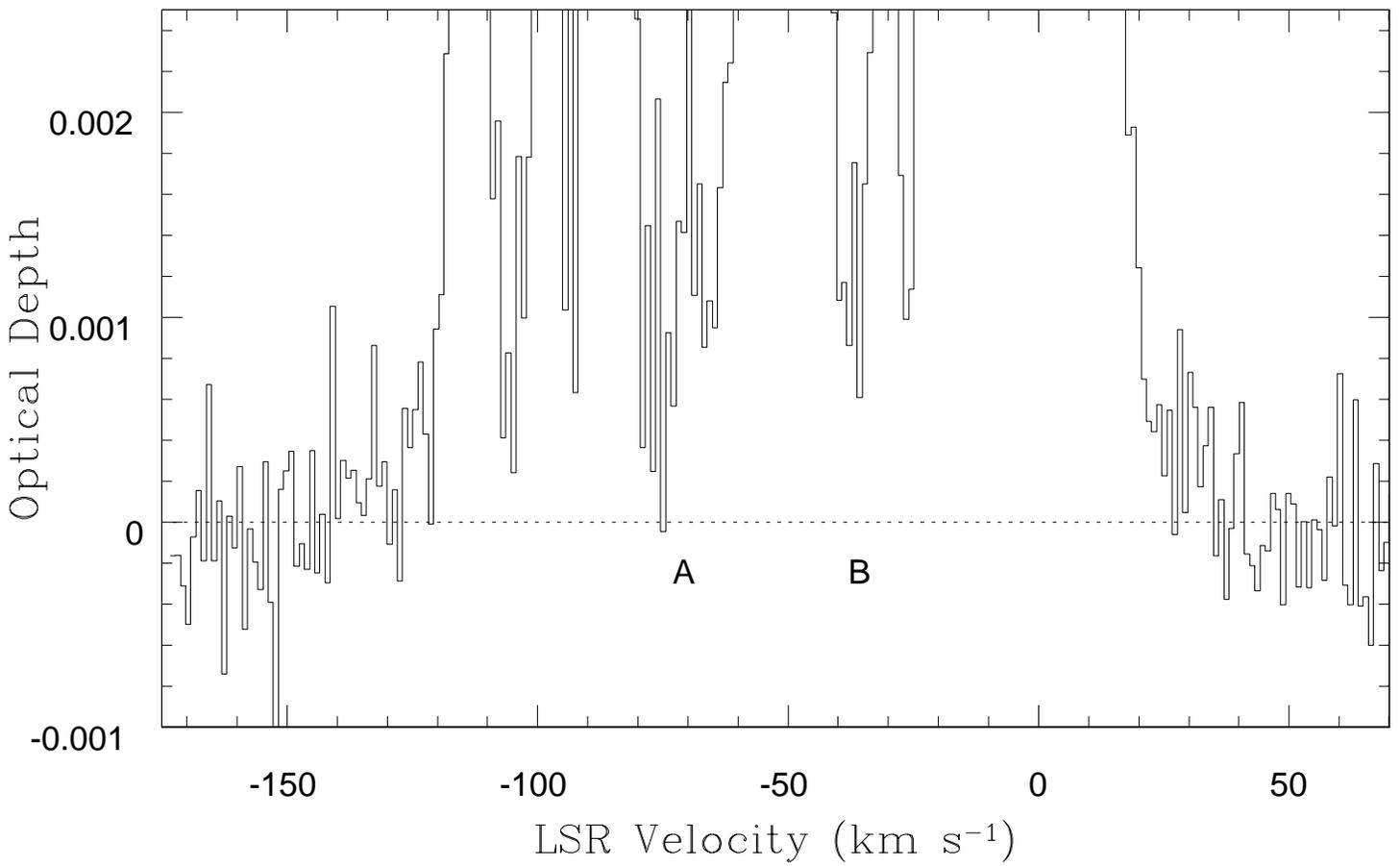
\end{document}